\documentclass[
reprint,
superscriptaddress,
amsmath,amssymb,
aps,
pra,
]{revtex4-2}

\usepackage{graphicx}
\usepackage{epstopdf}
\usepackage{xcolor}

\usepackage{bm}
\usepackage{gensymb}
\usepackage{siunitx}

\usepackage{array}
\usepackage{booktabs}
\usepackage{dcolumn}
\usepackage{tabularx}
\usepackage{multirow}

\usepackage{float}
\usepackage{subfig}

\usepackage{orcidlink}
\usepackage{tikz}
\usepackage{mwe}

\usepackage{textcomp}
\usepackage{verbatim}
\usepackage{url}

\newcolumntype{Y}{>{\centering\arraybackslash}X}
\newcolumntype{x}{D{.}{.}{6.6}}
\newcolumntype{y}{D{.}{.}{4.5}}
\newcolumntype{z}{D{.}{.}{5.7}}
\newcolumntype{f}{D{.}{.}{7.9}}
\newcolumntype{e}{D{.}{.}{5.6}}

\usepackage{hyperref}
\hypersetup{
    colorlinks=true,
    linkcolor=blue,
    anchorcolor=blue,
    citecolor=blue,
    urlcolor=blue
}

\begin{document}

\preprint{Radioactive Molecules}

\title{Towards Collinear Laser Spectroscopy of Radioactive Molecules Utilizing In-trap Produced Molecular Ion Beam}

\author{W.~C.~Mei\orcidlink{0009-0009-7635-6998}}
\affiliation{School of Physics and State Key Laboratory of Nuclear Physics and Technology, Peking University, Beijing 100871, China}

\author{S.~J.~Chen\orcidlink{0009-0005-1235-2411}}
\affiliation{School of Physics and State Key Laboratory of Nuclear Physics and Technology, Peking University, Beijing 100871, China}

\author{X.~F.~Yang\orcidlink{0000-0002-1633-4000}}
\email{xiaofei.yang@pku.edu.cn}
\affiliation{School of Physics and State Key Laboratory of Nuclear Physics and Technology, Peking University, Beijing 100871, China}

\author{J.~H.~Lv}
\affiliation{Institute of Modern Physics, Chinese Academy of Sciences, Lanzhou 730000, China }

\author{D.~Y.~Chen\orcidlink{0009-0006-7760-3338}}
\affiliation{School of Physics and State Key Laboratory of Nuclear Physics and Technology, Peking University, Beijing 100871, China}

\author{H.~R.~Hu\orcidlink{0009-0006-7760-3338}}
\affiliation{School of Physics and State Key Laboratory of Nuclear Physics and Technology, Peking University, Beijing 100871, China}

\author{Y.~F.~Guo\orcidlink{0009-0009-1817-7959}}
\affiliation{School of Physics and State Key Laboratory of Nuclear Physics and Technology, Peking University, Beijing 100871, China}

\author{Z.~Yan\orcidlink{0009-0007-8129-4600}}
\affiliation{School of Physics and State Key Laboratory of Nuclear Physics and Technology, Peking University, Beijing 100871, China}

\author{Y.~P.~Jing}
\affiliation{School of Physics and State Key Laboratory of Nuclear Physics and Technology, Peking University, Beijing 100871, China}

\author{C.~Zhang}
\affiliation{School of Physics and State Key Laboratory of Nuclear Physics and Technology, Peking University, Beijing 100871, China}

\author{Y.~P.~Lin}
\affiliation{School of Physics and State Key Laboratory of Nuclear Physics and Technology, Peking University, Beijing 100871, China}

\author{T.~X.~Gao\orcidlink{0009-0008-1025-6495}}
\affiliation{School of Physics and State Key Laboratory of Nuclear Physics and Technology, Peking University, Beijing 100871, China}

\author{X.~Shen}
\affiliation{School of Physics and State Key Laboratory of Nuclear Physics and Technology, Peking University, Beijing 100871, China}

\author{S.~W.~Bai\orcidlink{0000-0002-6087-9788}}
\affiliation{State Key Laboratory of Nuclear Physics and Technology, Institute of Quantum Matter, South China Normal University, Guangzhou 510006, China}

\author{R.~F.~Garcia Ruiz\orcidlink{0000-0002-2926-5569}}
\affiliation{Massachusetts Institute of Technology, Cambridge, Massachusetts 02139, USA}

\author{J.~Yang}
\affiliation{Institute of Modern Physics, Chinese Academy of Sciences, Lanzhou 730000, China }

\author{Y.~L.~Ye\orcidlink{0000-0001-8938-9152}}
\affiliation{School of Physics and State Key Laboratory of Nuclear Physics and Technology, Peking University, Beijing 100871, China}
\date{\today}      

\begin{abstract}
 Molecules containing short-lived isotopes, namely radioactive molecules, are among the most promising candidates for probing new physics beyond the Standard Model, although their production and spectroscopic measurements remain technically challenging. Here, we demonstrate an integrated methodology that combines formation of molecular ion beams in a radiofrequency quadrupole cooler-buncher with collinear laser spectroscopy. As a proof-of-principle experiment, we successfully produce molecular ions such as $\mathrm{BaF^{+}}$ and $\mathrm{YbF^{+}}$ via in-trap ion–molecule reactions and perform high-resolution laser spectroscopy of the target molecule $\mathrm{^{138}BaF}$. Vibrational and rotational structures of $\mathrm{^{138}BaF}$ across different electronic states are obtained using resonance-enhanced multiphoton ionization schemes, confirming the feasibility of the proposed methodology. This work establishes a practical route for future formation and spectroscopic studies of short-lived radioactive molecules, such as those containing $^{225}$Ra, at radioactive ion beam facilities.
\end{abstract}

\maketitle
\section{Introduction}
The origin of the imbalance between matter and antimatter in the universe remains an unsolved mystery that cannot be fully explained by the Standard Model~\cite{RevModPhys.76.1}. Additional sources of $CP$ violation from BSM (Beyond the Standard Model) physics are considered necessary to explain the matter-antimatter asymmetry~\cite{sakharov_violation_1991, RevModPhys.90.025008,RevModPhys.91.015001}. Precision measurements in atoms and molecules provide a perspective complementary to high-energy particle physics experiments to search for new physics beyond the Standard Model, and have increasingly emerged as a frontier field garnering widespread attention~\cite{RevModPhys.90.025008,RevModPhys.91.015001}.

In recent years, radioactive molecules containing octupole-deformed nucleus (e.g., $\mathrm{^{225}RaF}$, $\mathrm{^{225}RaOH^{+}}$, $\mathrm{^{227}AcF}$, etc.) have been regarded as promising candidates for measuring the $CP$-violating electric dipole moment (EDM)~\cite{Arrowsmith-Kron_2024}, due to the combination of the nuclear collective enhancement of their nuclear Schiff moment~\cite{PhysRevLett.76.4316,PhysRevC.56.1357} and the relatively large effective intra-molecular fields~\cite{PhysRevLett.19.1396}. However, the diverse and complex internal structures of molecules present significant challenges for their precise manipulation and spectroscopy studies. For example, techniques widely applied in precision measurements of cold atoms are not directly applicable to molecular systems. To achieve laser cooling of molecules, it is necessary to establish a quasi-closed cycling transition within their complex energy levels~\cite{DiRosa2004}. Consequently, a comprehensive spectroscopic survey of candidate molecules is essential to lay the foundation for designing experimental schemes for quantum-state preparation and readout, and ultimately for conducting precision measurements ~\cite{10.1063/1.3652333,GRAU201232,LOH201249,PhysRevA.90.062503,PhysRevA.91.042508,GRESH20161,ZHOU20191,PhysRevA.105.022823,mzd6-9kml}. 

However, studies of radioactive molecules are particularly challenging due to their low yields and short lifetimes. These limitations strongly motivate the development of advanced experimental techniques. For applications in future precision measurements of nuclear EDM, considerable experimental efforts have been devoted both to the production of radioactive molecules~\cite{AU2023375,AU2023144,Au2025,reiter2025,Charles_2022,PhysRevLett.126.023002,conn2025} and to systematic spectroscopic surveys of these species~\cite{GarciaRuiz2020,PhysRevLett.127.033001,Udrescu2024,PhysRevA.110.L010802,Athanasakis-Kaklamanakis2025,fvsy-v1q6,Wilkins2025,Athanasakis-Kaklamanakis2025AcF,conn2025}. In particular, the first successful measurement of radium monofluoride (RaF) at the CERN-ISOLDE radioactive ion beam (RIB) facility using high-sensitivity and high-resolution collinear resonance ionization spectroscopy initiated the field of laser spectroscopy of short-lived radioactive molecules~\cite{GarciaRuiz2020}.

To date, our limited understanding of the laser spectroscopy of radioactive molecules, especially those containing short-lived isotopes, has been primarily based on online experiments conducted at RIB facilities. Representative examples include RaF~\cite{GarciaRuiz2020,PhysRevLett.127.033001,Udrescu2024,PhysRevA.110.L010802,Athanasakis-Kaklamanakis2025,fvsy-v1q6,Wilkins2025} and AcF~\cite{Athanasakis-Kaklamanakis2025AcF}. For RaF$^+$ molecules, efficient production can be achieved by using a hot-cavity in-source method, which is a common approach for ion beam production at isotope separator online (ISOL)-type RIB facilities~\cite{EDER1992535,KIRCHNER1997135,Koster2007,KRONENBERG20084252}. By introducing reactive gasses (e.g., $\mathrm{CF_{4}}$, $\mathrm{SF_{6}}$) into the ISOL target and ion source, radioactive isotopes produced during proton irradiation can react to form molecules, which are subsequently ionized, extracted, mass separated, and delivered to the experimental terminals. Although high-temperature conditions of the target and ion source (typically around 2000~$^{\circ}$C) accelerate chemical reactions in kinetics, they can also lead to molecular dissociation for fragile species, such as polyatomic molecules, resulting in undesirably low yields. In addition, the high internal temperatures populate a large number of vibrational and rotational excited states, significantly complicating spectroscopic measurements and analysis.

Alternative approaches to produce cold molecules containing long-lived radioactive isotopes for precision measurements, which are commonly employed in the atomic, molecular and optical (AMO) physics community, include the use of offline ion traps~\cite{PhysRevLett.126.023002} or cryogenic molecular beam sources~\cite{conn2025}. These efforts have enabled the production and spectroscopic investigation of species such as $\mathrm{^{226}RaOH}$~\cite{conn2025}.
However, it is still a challenge to extend these methods to molecules containing short-lived isotopes produced directly at RIB facilities.

In experiments at RIB facilities, the formation of unintended radioactive molecules is frequently observed in radiofrequency quadrupole cooler-bunchers (RFQ-CB)~\cite{AU2023375} or gas-filled stopping cells~\cite{GREINER2020324}. For example, in ISOLTRAP RFQ-CB at CERN-ISOLDE, molecular formation through reactions between uranium ion beams and residual or buffer-gas contaminants are commonly observed~\cite{AU2023375}. Recently, by injecting $\mathrm{SF_{6}}$ into the RFQ beamline after the FRS Ion Catcher at GSI~\cite{reiter2025}, efficient production of radioactive molecules including $\mathrm{RaF^{+}}$ and $\mathrm{PbF^{+}}$ have been demonstrated. 

In this work, we demonstrate an integrated methodology that combines formation of molecular ion beams in a RFQ-CB with collinear laser spectroscopy, by carrying out systematic investigations using the PLASEN (Precision LAser Spectroscopy for Exotic Nuclei) experiment at Peking University~\cite{HU20252721}. Here, we report the detailed results of the production of $^{ A}$BaF$^+$ and $^{A}$YbF$^+$ molecules in an RFQ-CB, as well as the collinear laser spectroscopy measurement of $^{\rm 138}$BaF molecules. This work represents an important step towards future studies of short-lived radioactive molecules at RIB facilities in China, such as the Beijing Radioactive Ion-beam Facility (BRIF) ~\cite{NAN2025104188} and the High Intensity heavy-ion Accelerator Facility (HIAF)~\cite{Zhou2022}, and towards the online development of cold radioactive molecules using cryogenic RFQ-CB systems.

\begin{figure*}[htbp]
    \centering
    \includegraphics[width=0.98\textwidth]{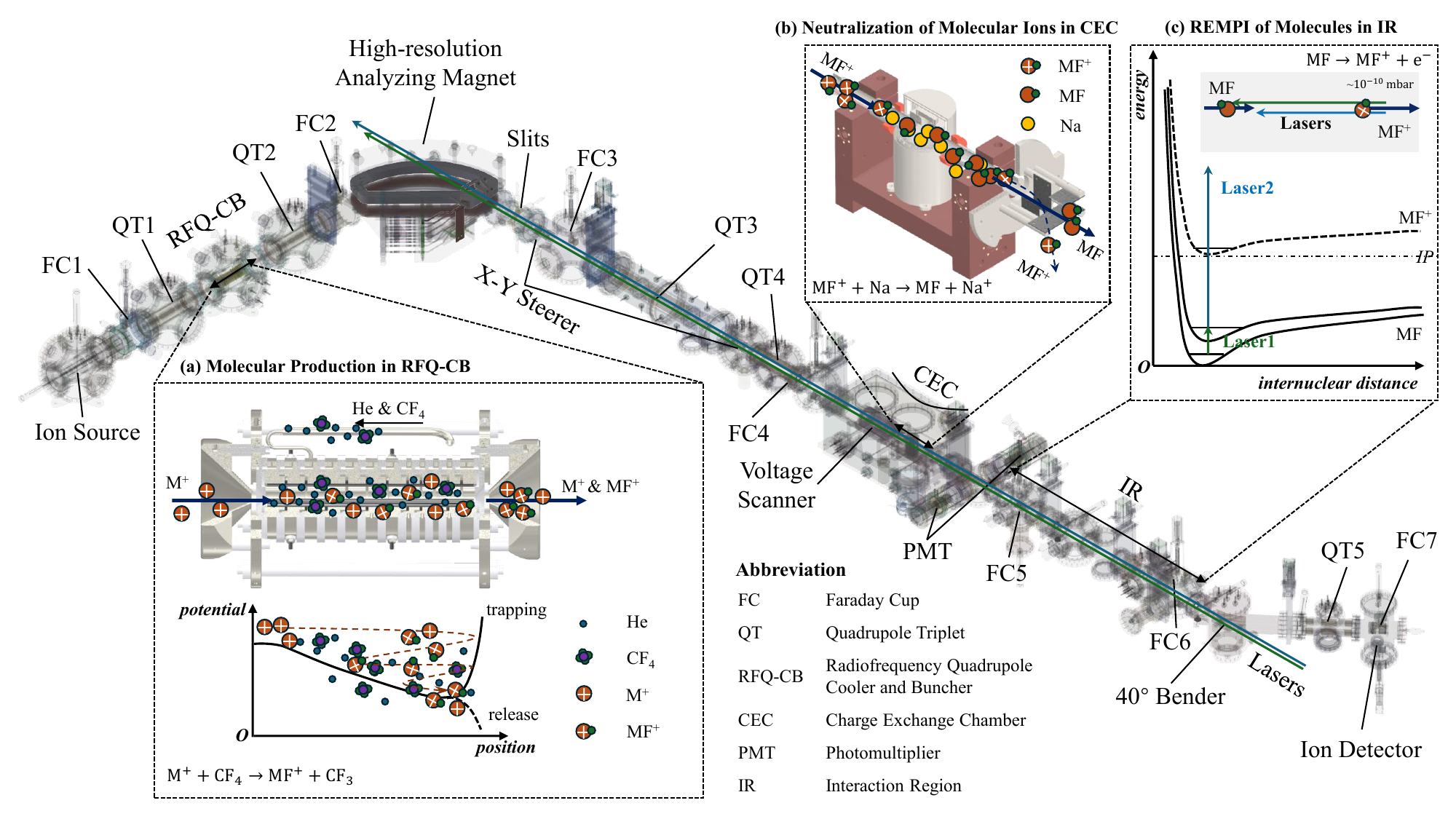}
    \captionsetup{justification=raggedright, singlelinecheck=false}
    \caption{Schematic of the offline PLASEN setup and their electrostatic components. The main systems required for performing collinear laser spectroscopy of molecules are highlithed:  (a) molecular production in RFQ-CB; (b) neutralization of molecular ions in CEC; (c) REMPI of molecules in IR.}
    \label{fig1}
\end{figure*}
\section{Experimental Setup}
As shown in Fig.~\ref{fig1}, the PLASEN system at Peking University consists of an ion source, an RFQ-CB, a mass-analyzing magnet, and a collinear laser spectroscopy setup, as described in detail in our previous publication~\cite{offlineCLS,ZHANG202337,HU20252721}. In the present work, the gas injection system of the RFQ-CB was modified to enable the production of molecular ion beams. An overview of the individual components in the experimental system is provided below.

First, in the ion source, a high-power 532 nm Nd:YAG laser ablates a metal target of Ba or Yb  at a repetition rate of 100 Hz to produce the corresponding ions. The ions are extracted by a set of Einzel lenses and subsequently accelerated to 30 keV, defined by the potential of the ion-source with respect to ground~\cite{offlineCLS}. The ion beam is then guided through the electrostatic ion beam components and injected into the RFQ-CB. To prevent an overload of ions in the RFQ-CB, the incoming ion beam is chopped by an electrostatic deflector synchronized with the laser ablation pulses. 

The ion beam is then injected into the RFQ-CB~\cite{Liu2025}. The RFQ-CB is floated on a high-voltage platform that is slightly lower than that of the ion source (approximately 29.98 kV), thereby decelerating the ions to less than 20 eV. The injected ions are temporarily confined within the RFQ-CB by the combined confining potentials of the radio-frequency quadrupole electrodes and axial DC electrodes. The RFQ-CB is typically filled with helium (He) buffer gas. Through collisions with the buffer-gas atoms, the ions gradually dissipate their kinetic energy until reaching thermal equilibrium with the He gas. To produce molecular ions within the RFQ-CB, a trace fraction of reactive gas $\mathrm{CF_4}$ can be introduced into the buffer gas, inducing ion–molecule reactions during the cooling process to form molecular species. To optimize the yields of the target molecular ions, the flow rates of the buffer gas and the reactive gas are independently regulated with high precision using dedicated mass flow controllers before being mixed in a defined proportion. After the reaction products have been accumulated for a certain period, the RFQ-CB releases all confined ions and re-accelerates them to approximately 29.98 keV for subsequent electromagnetic separation of reaction products and their isotopologues. A high-precision voltage divider, connected to the RFQ-CB chamber and read out by a 7.5-digit high-precision digital multimeter, continuously monitors and records the accurate beam energy in real time. All ionic molecular products, along with any remaining unreacted atomic ions, pass through the analyzing magnet for separation and purification. The magnet is designed with a resolving power exceeding 230, which is sufficient to distinguish different isotopologues of the same molecule. In addition, by scanning the magnetic field, mass spectrometric analysis of the ion–molecule reaction products generated within the RFQ-CB can be performed.

After being deflected by 90° in the dipole magnet, the molecular ions of interest enter the beamline for collinear laser spectroscopy. Along the beamline, various ion beam-optic elements, including quadrupole triplet (QT) lenses and steering electrodes (X-Y Steerer), are used to adjust the beam trajectory and focus, ensuring transmission efficiency and sufficient overlap with collinear laser beams. Multiple Faraday cups (FC1–FC7 shown in Fig.~\ref{fig1}) are installed for beam diagnostics and beam tuning~\cite{Yanzhou2025}. Since our primary interest lies in neutral molecules in this work, the molecular ion beam must be neutralized before interacting with the lasers. This is achieved using a charge exchange cell (CEC)~\cite{onlineCLS}, which contains an oven that can be loaded with an alkali metal such as sodium (Na). When heated, the alkali metal vaporizes and fills the interior of the CEC. As the molecular ions pass through the cell, they capture an electron from the outer shell of the alkali atoms and are converted into neutral species. Downstream of the charge exchange cell, a biased electrostatic deflector is used to remove any remaining molecular ions that fail to neutralize, while the neutral molecular beam continues towards the interaction region (IR).

In the interaction region, the molecules encounter laser pulses that are time-synchronized with the molecular bunches. When the laser frequency satisfies the resonance condition, the molecules are excited and ionized sequentially by the lasers. Since the molecular beam propagates collinearly with the laser and has a kinetic energy of $\sim$30 keV, Doppler broadening is suppressed to several tens of MHz due to the narrowed velocity distribution during acceleration, making the setup suitable for resolving the rotational and hyperfine structures of certain molecules. The molecular ions produced in the resonance-enhanced multiphoton ionization (REMPI) process are guided by an electrostatic bender after the interaction region and detected by an ion detector at the end of the beamline, providing real-time counting-rate information of the molecular ions. The raw signals from the ion detector are amplified by a fast amplifier and processed by a constant-fraction discriminator (CFD) before being sent to a time-to-digital converter (TDC). The TDC records the time of flight (TOF) from the release of the molecular ions by the RFQ-CB to their arrival at the detector. By statistically analyzing and filtering the TOF spectrum event by event, some noise signals can be effectively excluded. During laser frequency scanning, the high voltage of the RFQ-CB (which determines the molecular beam energy), the laser frequency measured by high-precision wavemeter, and the ion-event information are recorded and stored by a self-developed data acquisition (DAQ) system~\cite{Liu2023}. The DAQ system can also provide real-time reconstruction of the measured laser spectrum and TOF spectrum.

Since the electronic, vibrational, and rotational motions of molecules exhibit distinct energy-level structures, they impose different requirements on spectroscopic resolution. Meanwhile, the laser scheme used in the REMPI measurements may involve multiple laser beams with different wavelengths. Therefore, the experimental system is equipped with various types of lasers to meet the requirements for different wavelengths, linewidths, and power levels. In experiments targeting the BaF molecules, a broadband pulsed Ti:Sapphire (Ti:Sa) laser and a pulsed optical parametric oscillator (OPO) are used to excite the first-step or intermediate transitions in the REMPI process. In addition, a Nd:YAG laser provides high-power 1064 nm pulses for non-resonant direct ionization of the molecules.

\section{In-trap Molecular Production}
In the experimental setup shown in Fig.~\ref{fig1}, the RFQ-CB essentially functions as a linear Paul ion trap that accumulates confined ions and releases them periodically. Within the ion trap, it is commonly observed that trapped ions collide and react with active gas components to form molecular species~\cite{PhysRevA.101.022705,PhysRevLett.126.023002,PhysRevA.62.011401,Xu08012026}.
\begin{figure}[htbp]
    \centering
    \includegraphics[width=0.97\columnwidth]{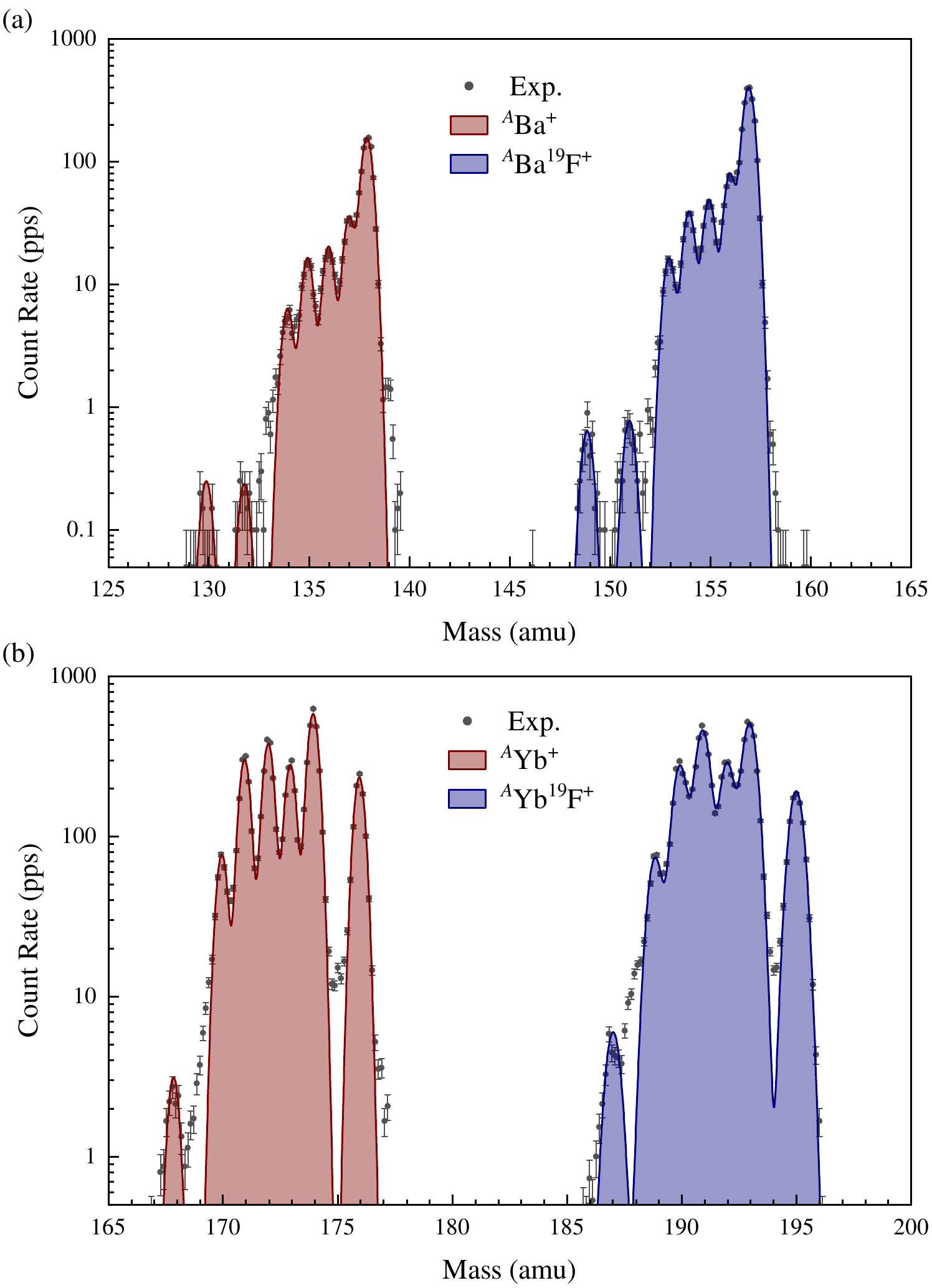}
    \captionsetup{justification=raggedright, singlelinecheck=false}
    \caption{Mass spectra of the molecular products and remaining unreacted reactants from ion-molecule reactions in the RFQ-CB: (a) $\mathrm{Ba^{+}+CF_4}$; (b) $\mathrm{Yb^{+}+CF_4}$}
    \label{fig2}
\end{figure}

Although gas-phase ion–molecule reactions have been extensively studied in chemistry for a wide range of elements and species~\cite{bowers1979gas}, experimental data on reactions that produce radioactive molecular candidates for EDM experiments, such as RaF and RaOH, remain relatively scarce. The production of these target radioactive molecules via in-trap ion–molecule reactions still requires systematic experimental investigation~\cite{AU2023375,Charles_2022,reiter2025}. Therefore, in this work, we use the alkaline-earth element Ba and the alkaline-earth-like element Yb, which share similar chemical properties to Ra, to investigate and validate the feasibility of producing  molecular ions $\mathrm{BaF^{+}}$ and $\mathrm{YbF^{+}}$ in the RFQ-CB.

With a trace amount of $\mathrm{CF_{4}}$ injected into the RFQ-CB, molecular ions $\mathrm{BaF^{+}}$ and $\mathrm{YbF^{+}}$ are produced, as shown in the mass spectra (Fig.~\ref{fig2}) obtained by counting the ions as a function of the current applied to the analyzing magnet. By adjusting the current of the analyzing magnet, different reaction products generated in the RFQ-CB can be selectively separated from the original atomic ion beams, enabling the transmission of purified molecular ion beams. In addition, molecular ions containing various natural isotopes of Ba or Yb, including stable and long-lived ones, are clearly identified in the spectra. 

\begin{figure}[htbp]
    \centering
    \includegraphics[width=0.97\columnwidth]{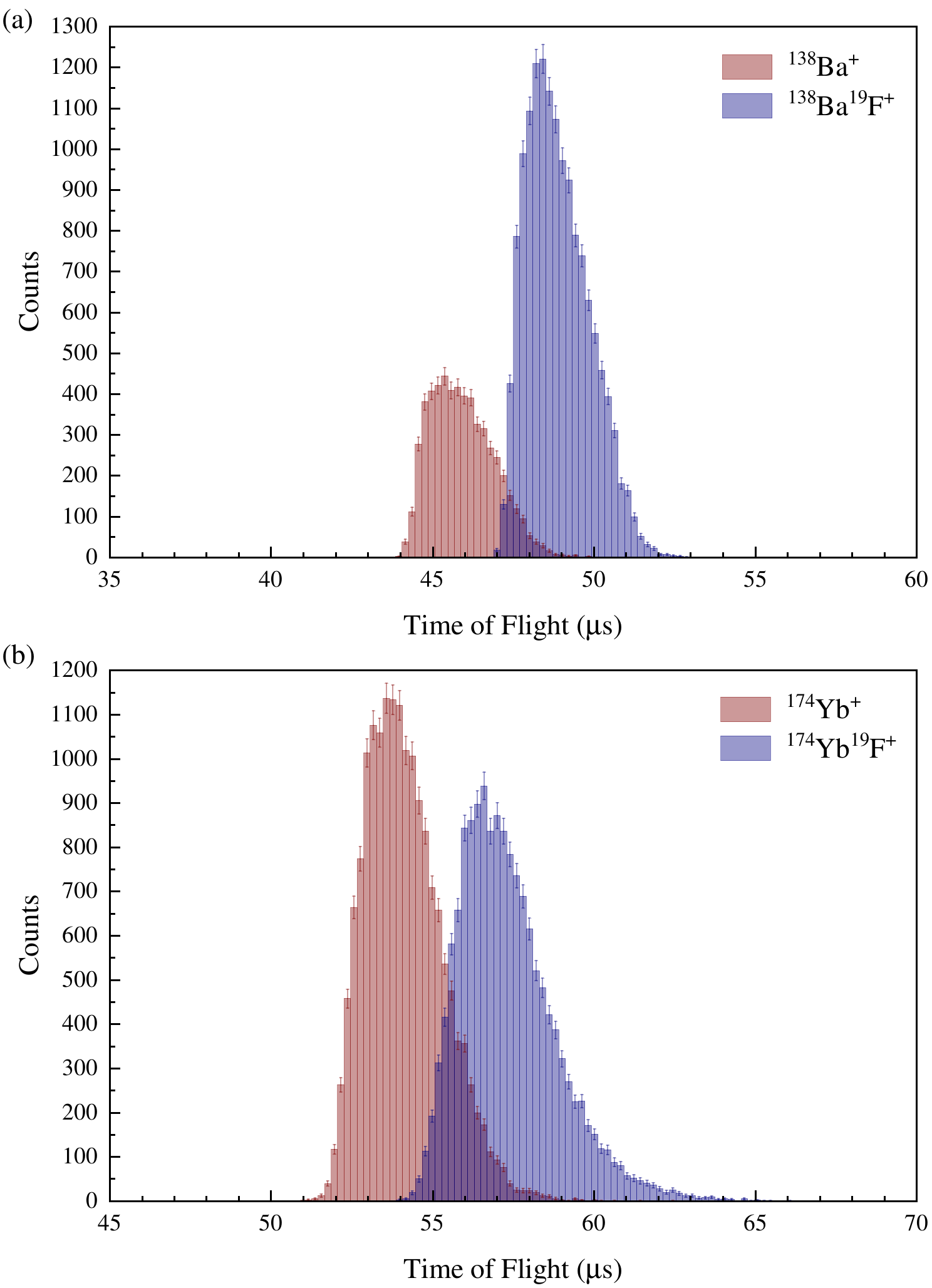}
    \captionsetup{justification=raggedright, singlelinecheck=false}
    \caption{Time of flight (TOF) spectra of the molecular products and residual reactants from ion-molecule reactions in the RFQ-CB: (a) $\mathrm{Ba^{+}+CF_4}$; (b) $\mathrm{Yb^{+}+CF_4}$}
    \label{fig3}
\end{figure}

The molecular ion beams extracted from the RFQ-CB are pulsed, with a temporal width of less than \SI{3}{\micro\second} FWHM (Full Width at Half Maximum) in the time-of-flight (TOF) spectrum, as shown in Fig.~\ref{fig3}. This corresponds to a bunch with a spatial length of approximately \SI{0.5}{\meter}, considering the beam energy of about \SI{30}{\kilo\electronvolt}. The bunch length is comparable to the spatial scale of the interaction region, allowing efficient interaction between the molecules and the laser beams. The pulse width of the extracted molecular ion beam can be further compressed by adjusting the DC trapping potential but at the expense of reduced transmission efficiency through the RFQ-CB~\cite{Liu2025}.

\begin{figure}[htbp]
    \centering
    \includegraphics[width=0.97\columnwidth]{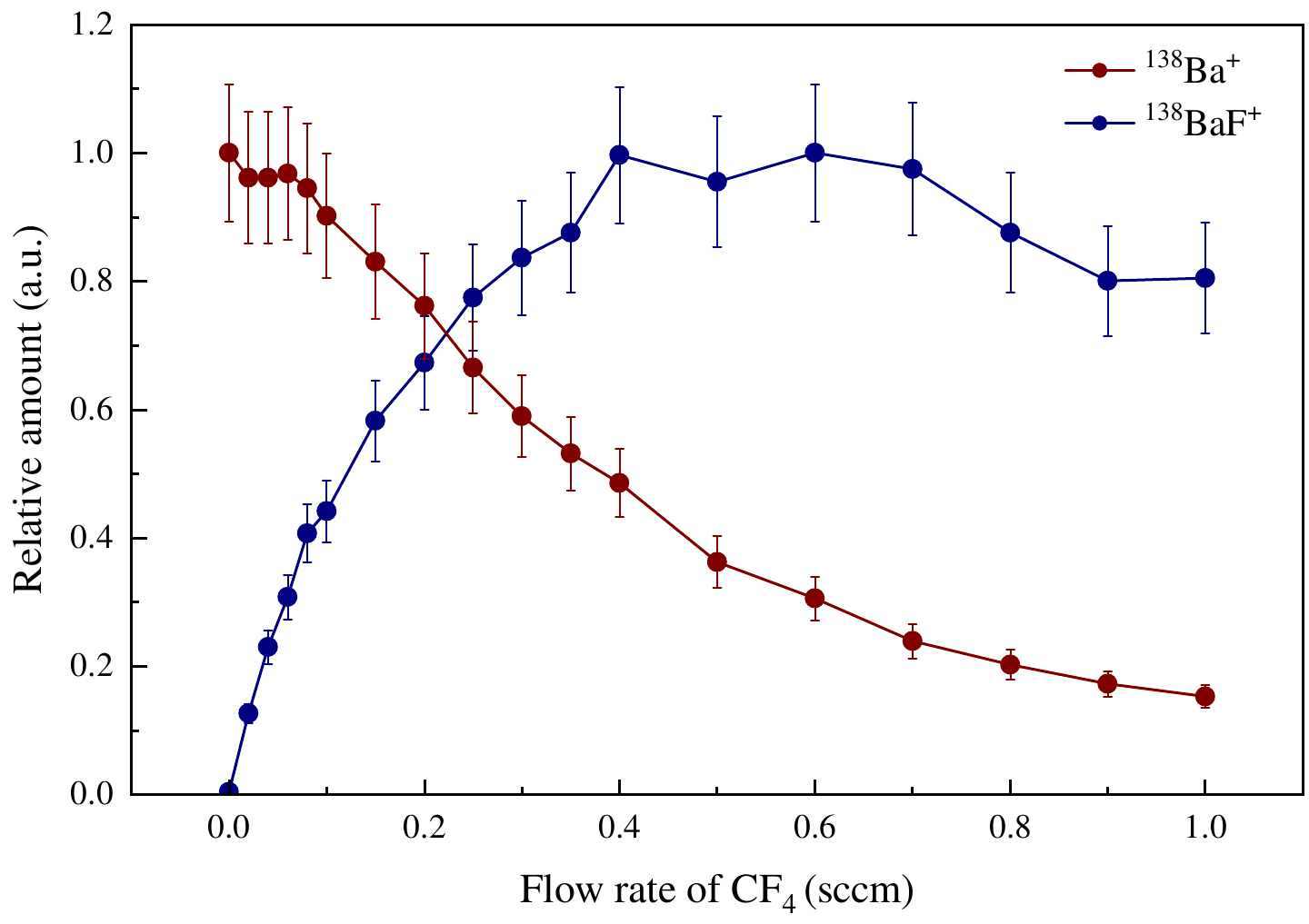}
    \captionsetup{justification=raggedright, singlelinecheck=false}
    \caption{Relative amount of formed molecular ions $\mathrm{BaF^{+}}$ created in the RFQ-CB and the residual amount of reactants $\mathrm{Ba^{+}}$ as a function of the flow rate used for $\mathrm{CF_{4}}$. The relative amount of $\mathrm{Ba^{+}}$ and $\mathrm{BaF^{+}}$ indicated on the y-axis are normalized to their respective maximum values and are given in arbitrary units (a.u.).}
    \label{fig4}
\end{figure}

By controlling the reaction conditions, the in-trap production efficiency of molecular ions can be optimized. As shown in Fig.~\ref{fig4}, the yield of $\mathrm{BaF^{+}}$ increases gradually with increasing $\mathrm{CF_{4}}$ flow rate until it reaches saturation at approximately 0.4 sccm. Although the $\mathrm{BaF^{+}}$ yield saturates beyond this point, a further increase in $\mathrm{CF_{4}}$ flow rate leads to a gradual decrease in the remaining $\mathrm{Ba^{+}}$ ions, suggesting the formation of additional by-products and the loss of reactant $\mathrm{Ba^{+}}$ ions.
\begin{table*}[htbp]
    \centering
    \captionsetup{justification=raggedright, singlelinecheck=false}
    \caption{Calculated enthalpy changes ($\Delta H$) for ion–molecule reactions $\mathrm{M^{+}}+\mathrm{CF_{4}}\rightarrow \mathrm{MF^{+}}+\mathrm{CF_{3}}$ and $\mathrm{M^{+}}+\mathrm{SF_{6}}\rightarrow \mathrm{MF^{+}}+\mathrm{SF_{5}}$, where $\mathrm{M}=\mathrm{Mg},~\mathrm{Ca},~\mathrm{Sr},~\mathrm{Ba},~\mathrm{Ra},~\mathrm{Yb}$. The values are derived from thermochemistry calculations ($\Delta H_1$) and ionization potentials of M atoms ($\Delta H_2$)~\cite{kramida2024nist} and MF molecules ($\Delta H_3$)~\cite{MgF1967,PhysRevA.109.022813,PhysRevA.48.3012,doi:10.1021/acs.jpca.5c03336,fvsy-v1q6}.}
    \label{table1}
    \setlength{\tabcolsep}{11pt}
    \setlength{\extrarowheight}{3pt}
    \begin{tabular}{c c c c c c c}
        \toprule
        Reactive Gas & Element & Valence Electron & {$\Delta H_1$ (\si{eV})} & {$\Delta H_2$ (\si{eV})} & {$\Delta H_3$ (\si{eV})} & {$\Delta H$ (\si{eV})} \\
        \midrule
        \multirow{6}{*}{CF\textsubscript{4}} & Mg & $3s^2$ & +0.82 & 7.65 & 7.68 & +0.85 \\
        & Ca & $4s^2$ & +0.14 & 6.11 & 5.83 & -0.15 \\
        & Sr & $5s^2$ & +0.04 & 5.69 & 5.36$\sim$5.62 & -0.29$\sim$-0.03 \\
        & Ba & $6s^2$ & -0.43 & 5.21 & 4.80 & -0.84 \\
        & Ra & $7s^2$ & -0.47 & 5.27 & 4.97 & -0.78 \\
        & Yb & $4f^{14}6s^2$ & +0.78 & 6.25 & 6.04 & +0.57 \\
        \midrule
        \multirow{6}{*}{SF\textsubscript{6}} & Mg & $3s^2$ & -0.74 & 7.65 & 7.68 & -0.71 \\
        & Ca & $4s^2$ & -1.43 & 6.11 & 5.83 & -1.71 \\
        & Sr & $5s^2$ & -1.52 & 5.69 & 5.36$\sim$5.62 & -1.85$\sim$-1.59 \\
        & Ba & $6s^2$ & -1.99 & 5.21 & 4.80 & -2.40 \\
        & Ra & $7s^2$ & -2.03 & 5.27 & 4.97 & -2.34 \\
        & Yb & $4f^{14}6s^2$ & -0.78 & 6.25 & 6.04 & -1.00 \\
        \bottomrule
    \end{tabular}
\end{table*}

In the RFQ-CB, reactant ions may be lost or converted into undesired by-products, making it difficult to directly determine the intrinsic reaction conversion rate with $\mathrm{CF_{4}}$. To quantitatively evaluate the overall performance of the in-trap molecular production, we treat the RFQ-CB as a black-box converter, without explicitly considering the detailed internal reaction pathways. We define the overall conversion efficiency as the ratio of the extracted molecular ion beam current (measured by FC3 after the analyzing magnet) to the injected atomic ion beam current (measured by FC1 immediately after the ion source). After optimization, the overall efficiency for producing $\mathrm{BaF^{+}}$ and $\mathrm{YbF^{+}}$ both exceeds 10\%. This value is significant because it includes not only the intrinsic reaction conversion efficiency, but also the cooling and transmission efficiency through the RFQ-CB, and along the beamline up to FC3. 

It should be noted that the original ion beam generated by the laser ablation ion source exhibits relatively poor beam quality. This is reflected in an overall cooling and transmission efficiency of approximately 30\% for $\mathrm{Ba^{+}}$ or $\mathrm{Yb^{+}}$ beams passing through the RFQ-CB. In contrast, alkali or alkaline-earth ion beams produced from the surface ion source, exhibit significantly better beam quality, and thus will lead to significantly improved RFQ-CB transmission efficiency. For example, in a separate experiment, the measured cooling and transmission efficiency of the RFQ-CB exceeded 70\% for a $\mathrm{Rb^{+}}$ beam produced from a surface ion source~\cite{Liu2025,HU20252721}. A high-quality beam from the surface ion source would substantially enhance the overall efficiency of $\mathrm{BaF^{+}}$ and $\mathrm{YbF^{+}}$ production via the in-trap method.

The successful production of $\mathrm{BaF^{+}}$ and $\mathrm{YbF^{+}}$ molecules in the RFQ-CB opens new possibilities for the in-trap production of other molecular species, particularly radioactive molecules such as $\mathrm{RaF^{+}}$, which is one of our primary targets for fundamental physics research. To explore the feasibility of producing other molecular species in the RFQ-CB, we performed thermochemical calculations for the ion–molecule reactions
$\mathrm{M^{+}}+\mathrm{CF_{4}}\rightarrow \mathrm{MF^{+}}+\mathrm{CF_{3}}$ and
$\mathrm{M^{+}}+\mathrm{SF_{6}}\rightarrow \mathrm{MF^{+}}+\mathrm{SF_{5}}$,
with $\mathrm{M}=\mathrm{Mg},~\mathrm{Ca},~\mathrm{Sr},~\mathrm{Ba},~\mathrm{Ra},~\mathrm{Yb}$, as summarized in Table~\ref{table1}.

The reaction enthalpies in Table~\ref{table1} are derived from thermodynamic data for gas-phase reactions at approximately 300~K, which are obtained using HSC Chemistry 9 (v.9.8.1.2, Outotec, 2018). For example, the reaction
$\mathrm{M} (g) + \mathrm{CF_4} (g) \rightarrow \mathrm{MF} (g) + \mathrm{CF_3} (g)$
may have an enthalpy change $\Delta H_1$ according to HSC Chemistry. By referring to the ionization potential ($IP$) of the atom $\mathrm{M}$
\begin{equation}
\mathrm{M} (g) \rightarrow \mathrm{M^{+}} (g) + e^{-} ,~\Delta H_2=+IP(\mathrm{M})
\end{equation}
and of the molecule $\mathrm{MF}$,
\begin{equation}
\mathrm{MF} (g) \rightarrow \mathrm{MF^{+}} (g) + e^{-},~\Delta H_3=+IP(\mathrm{MF})
\end{equation}
we can obtain the overall reaction enthalpy $\Delta H=\Delta H_1 -\Delta H_2+\Delta H_3$. 

Except for the reactions $\mathrm{Mg^{+}}+\mathrm{CF_{4}}$ and $\mathrm{Yb^{+}}+\mathrm{CF_{4}}$, all other ion–molecule reactions calculated are exothermic ($\Delta H<0$), indicating a strong thermodynamic tendency towards spontaneous reaction. Even for endothermic reactions, the kinetic energy of the ions within the RFQ-CB may be sufficient to overcome the reaction barrier, as demonstrated by the production of $\mathrm{YbF}^+$ molecules in Fig.~\ref{fig2}~(b). Therefore, the production of radioactive $\mathrm{RaF^{+}}$ via the in-trap method appears to be highly promising in thermodynamics, as the reaction enthalpy for forming $\mathrm{RaF}^+$ is comparable to that for forming $\mathrm{BaF}^+$. Furthermore, thermodynamic calculations suggest that $\mathrm{SF_6}$ may be a more favorable reactive gas than $\mathrm{CF_4}$. The influence of different reactant gasses on the conversion efficiency for molecular ion production in the RFQ-CB requires further systematic experimental investigation.

\section{Collinear Resonance Ionization Spectroscopy of In-trap Produced Molecules}
The molecular ion beam produced in the RFQ-CB is neutralized via charge-exchange reactions with Na in the CEC and can subsequently be used for collinear laser spectroscopy experiments based on either laser-induced fluorescence (LIF) or resonance-enhanced multiphoton ionization (REMPI) detection. Among these methods, REMPI detection offers higher sensitivity due to its low background, and it has been widely employed in laser spectroscopy of atoms and molecules involving short-lived and low-yield unstable nuclei~\cite{YANG2023104005}. When implemented in a collinear geometry for high resolution, this technique is commonly referred to as collinear resonance ionization spectroscopy, which combines the advantages of high spectral resolution and high detection sensitivity~\cite{YANG2023104005}.

Consequently, we have carried out a proof-of-principle collinear resonance ionization spectroscopy experiment on $\mathrm{^{138}BaF}$ molecules using the in-trap produced molecular ion beams. To investigate the various internal degrees of freedom of $\mathrm{^{138}BaF}$, including vibrational and rotational structures, lasers with different linewidths are used and distinct REMPI schemes are employed, as described in the following sections.

\subsection{Vibrational Structure of $\mathrm{^{138}BaF}$ molecules}
\begin{figure}[htbp]
    \centering
    \includegraphics[width=0.97\columnwidth]{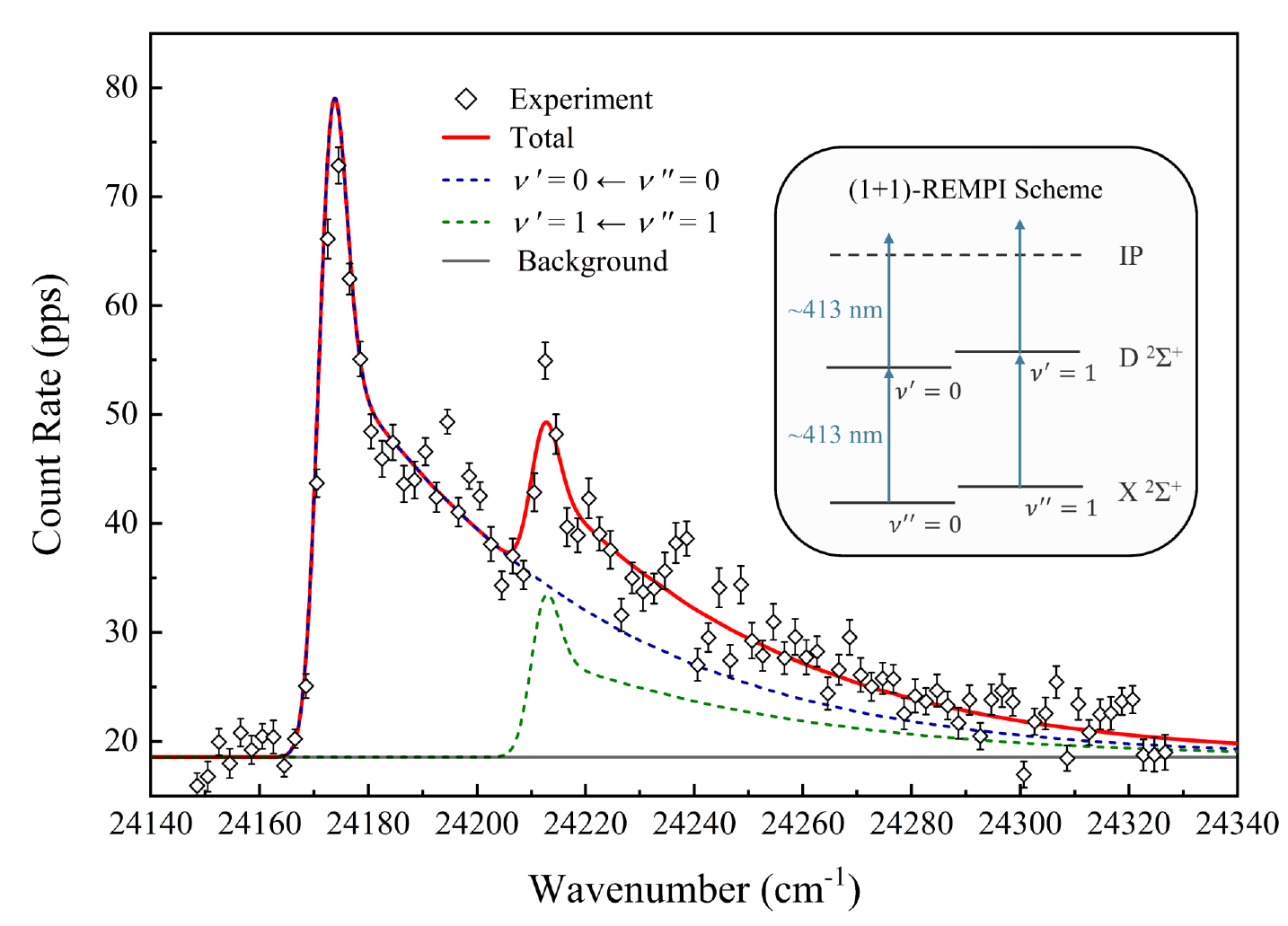}
    \captionsetup{justification=raggedright, singlelinecheck=false}
    \caption{Vibronic spectrum of $\mathrm{D~^{2}\Sigma^{+}\leftarrow X~^{2}\Sigma^{+}}$ transition in $\mathrm{^{138}BaF}$ through (1+1)-REMPI scheme (inset).}
    \label{fig5}
\end{figure}

To investigate the vibrational structure of the $\mathrm{^{138}BaF}$ molecules, a monochromatic resonance-enhanced two-photon ionization scheme is employed, as illustrated in the inset of Fig.~\ref{fig5}. In this measurement, $\mathrm{^{138}BaF}$ molecules are excited via the $\mathrm{D~^{2}\Sigma^{+}\leftarrow X~^{2}\Sigma^{+}}$ transition using a 413-nm laser generated by a broadband OPO with a linewidth of several $\mathrm{cm^{-1}}$. Subsequent ionization occurs through the absorption of a second 413-nm photon, forming a (1+1)-REMPI scheme.

By scanning the laser frequency, we obtain the vibronic spectrum of the $\mathrm{D~^{2}\Sigma^{+}\leftarrow X~^{2}\Sigma^{+}}$ transition, as shown in Fig.~\ref{fig5}. The spectrum exhibits two prominent narrow peaks, each accompanied by a trailing structure on its right side (higher-frequency). Based on simulations using PGOPHER~\cite{WESTERN2017221} and molecular constants reported in the literature~\cite{PhysRevA.108.062812,BERNARD1992174,Effantin10081990}, these two narrow peaks are assigned to the band heads formed by the $P$-branch of the (0,0) and (1,1) bands, respectively, where $(\nu',\nu'')$ denotes the vibrational transition $\nu' \leftarrow \nu''$. The trailing structures on the higher-frequency side arise from the unresolved rotational bands of the $P$- and $R$-branch transitions.

Due to the broad linewidth of the OPO laser, the individual rotational transitions cannot be resolved in the measured spectrum. Therefore, a contour-fitting analysis was performed using PGOPHER by fixing the molecular constants of the ground state $\mathrm{X}~^{2}\Sigma^{+}$~\cite{PhysRevA.108.062812} and the excited state $\mathrm{D}~^{2}\Sigma^{+}$~\cite{BERNARD1992174,Effantin10081990} to the values reported in the literature. As a result of the charge-exchange process between the molecular ions and the high-temperature sodium vapor in the CEC, multiple vibrational and rotational states are populated in neutral $\mathrm{^{138}BaF}$ molecules. From the contour-fitting results in Fig.~\ref{fig5}, we infer that rovibrational excited states in the $\mathrm{X}~^{2}\Sigma^{+}$ state of $\mathrm{BaF}$ molecules characterized by $J'' \lesssim 100$ and $\nu'' \leq 1$ are at least efficiently populated.

To evaluate the overall detection efficiency, we measured the $\mathrm{^{138}BaF^{+}}$ beam intensity after magnetic separation, as well as the resonant ionization signal and background count rates when the laser frequency was tuned to the maximum in the spectrum. After optimization, the detection efficiency, defined as the ratio of detected resonant ion counts to the intensity of incident molecular beams, reaches approximately 0.01\% using the present scheme and experimental conditions. 

Considering that a broadband laser is used to simultaneously excite multiple rotational transitions, this efficiency is relatively modest for the collinear resonance ionization spectroscopy technique. On the one hand, the available laser power for both molecular excitation and ionization is significantly below the typical saturation power required for (1+1)-REMPI~\cite{D0CP03583A}, being limited by the maximum output energy (approximately 4.5 mJ per pulse) of the OPO laser employed in the experiment. Increasing the laser power is therefore expected to substantially enhance the detection efficiency. On the other hand, during the neutralization process in the CEC, molecular ions may populate not only the electronic ground state but also several low-lying excited electronic states, thereby reducing the effective population available for resonant excitation. In particular, neutral molecules in long-lived metastable states, such as the $\mathrm{A'}~^{2}\Delta$ state in BaF~\cite{BARROW1988535}, act as dark states in the measurement and contribute only to the background rather than to the signal of resonant ionization.

\subsection{Rotational Structure of $\mathrm{^{138}BaF}$ molecules}
Since Doppler broadening is substantially suppressed in the collinear laser spectroscopy geometry to a level comparable to natural broadening of the transition, the spectral resolution in the present experiment is primarily limited by the laser linewidth. To achieve rotational-structure resolved spectroscopy, a high-precision laser with a narrower linewidth is therefore required to selectively excite individual rotational transitions. 

\begin{figure}[htbp]
    \centering
    \includegraphics[width=0.97\columnwidth]{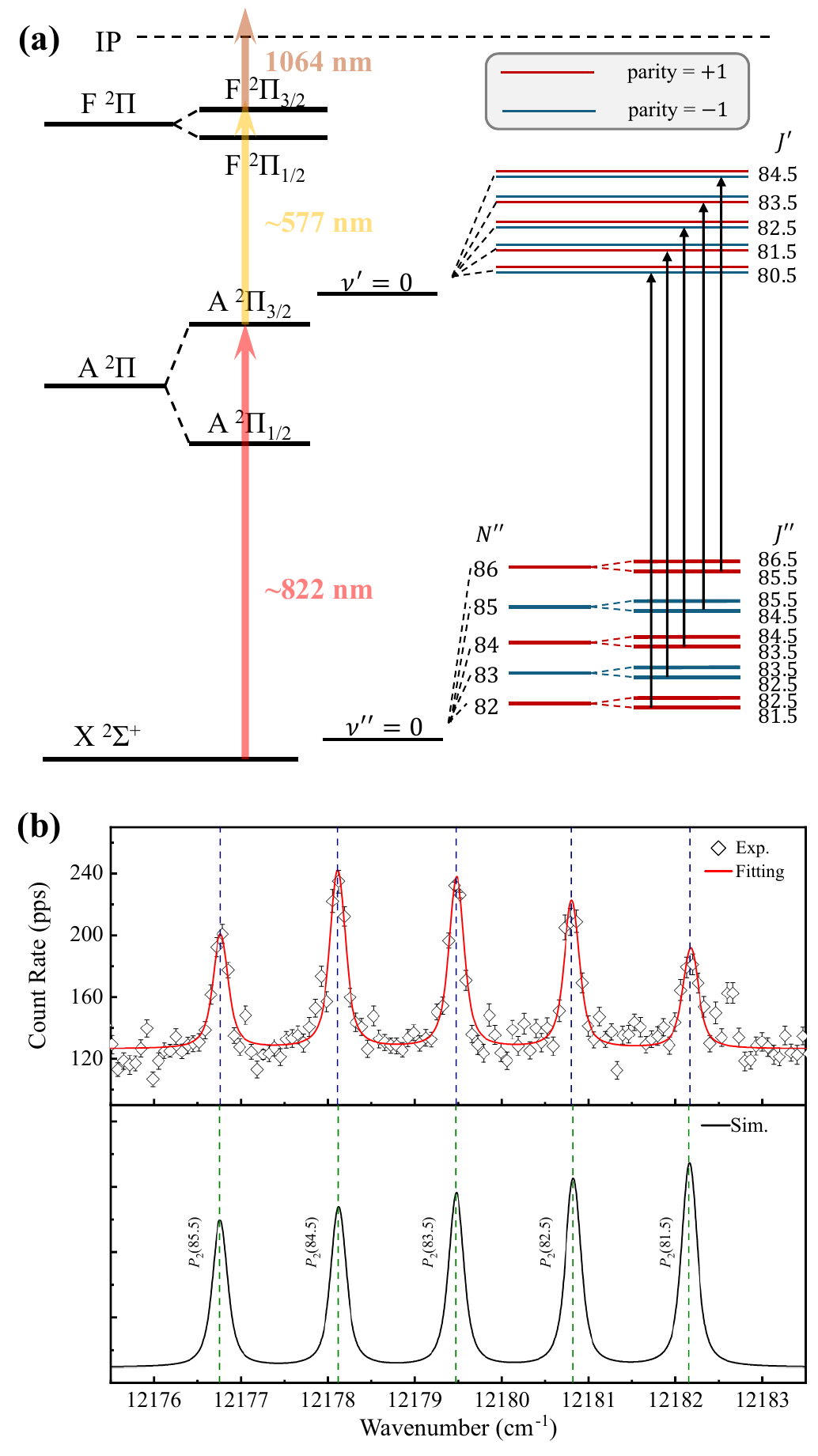}
    \captionsetup{justification=raggedright, singlelinecheck=false}
    \caption{(a)~(1+$1'$+$1''$)-REMPI scheme for measuring rotational structure-resolved spectrum of $\mathrm{A~^{2}\Pi_{3/2}\leftarrow X~^{2}\Sigma^{+}}~(0,0)$ transition in $\mathrm{^{138}BaF}$ molecules. (b)~Rotationally-resolved spectrum of $\mathrm{A~^{2}\Pi_{3/2}\leftarrow X~^{2}\Sigma^{+}} (0,0)$ transition in $\mathrm{^{138}BaF}$ molecules.}
    \label{fig6}
\end{figure}

For this purpose, we employed a ($1$+$1'$+$1''$)-REMPI scheme involving three sequential laser pulses, as illustrated in Fig.~\ref{fig6}~(a). A pulsed Ti:Sa laser with a linewidth of a few $\mathrm{GHz}$ is used to excite the $\mathrm{A~^{2}\Pi_{3/2}\leftarrow X~^{2}\Sigma^{+}}~(0,0)$ transition. Subsequently, an OPO laser operating near 577 nm drives the intermediate transition $\mathrm{F~^{2}\Pi_{3/2}\leftarrow A~^{2}\Pi_{3/2}}~(0,0)$. Finally, molecules excited to the high-lying $\mathrm{F~^{2}\Pi_{3/2}}$ state are ionized by a high-power 1064-nm pulse from a Nd:YAG laser.

Due to the transition selection rules and the limited linewidth of the intermediate laser, the accessible rotational branches and the range of rotational quantum numbers $J$ observed in the $\mathrm{A~^{2}\Pi_{3/2}\leftarrow X~^{2}\Sigma^{+}}$ spectrum are determined by the wavelength of the OPO laser. Figure~\ref{fig6}~(b) presents five measured $P$-branch rotational lines obtained with the OPO wavelength fixed at 577.40 nm (in air).

By comparing with simulations based on molecular constants reported in the literature~\cite{PhysRevA.108.062812,BERNARD1992174,Effantin10081990}, these five lines are assigned to the transitions $P_2(81.5)\sim P_2(85.5)$. We adopt the notation $\Delta J_{a,b}(J'')$ to label the rotational transitions, where $\Delta J = J' - J''$, and the primed and double-primed quantum numbers refer to the excited and ground states, respectively. The index $a$ equals 1 for transitions to the $\mathrm{A~^{2}\Pi_{1/2}}$ state, and 2 for transitions to the $\mathrm{A~^{2}\Pi_{3/2}}$ state. To specify which spin–rotation component of the ground state $\mathrm{X~^{2}\Sigma^{+}}$ participates in the transition, the index $b$ takes the value 1 for transitions originating from $J'' = N'' + 1/2$ and 2 for those from $J'' = N'' - 1/2$. When $a = b$, the redundant subscript is omitted. As usual, the labels $P$, $Q$, and $R$ denote transitions with $\Delta J = -1, 0,$ and $+1$, respectively. The assignments corresponding to the five experimentally observed spectral lines are marked in Fig.~\ref{fig6}~(b).

Although each rotational level of the excited $\mathrm{A~^{2}\Pi_{3/2}}$ state may be split into a $\Lambda$-doublet, parity selection rules restrict the allowed transitions. In Fig.~\ref{fig6}~(a), energy levels with parity $+1$ are marked with red color, while the parity $-1$ levels are marked with blue color. Consequently, all spectral lines observed in Fig.~\ref{fig6}~(b) correspond to transitions populating the lower component of the $\Lambda$-doublet in the excited state, as illustrated in Fig.~\ref{fig6}~(a). 

For data analysis, the observed spectral profile in Fig.~\ref{fig6}~(b) is modeled as a superposition of five Voigt functions
$V_j(\tilde{\nu}; \tilde{\nu}_{0,j}, A_j, \sigma, \Gamma)$, with $j=1\sim5$, on top of a constant background $C$, expressed as
\begin{equation}
F(\tilde{\nu}) = C + \sum_{j=1}^{5} V_j(\tilde{\nu}; \tilde{\nu}_{0,j}, A_j, \sigma, \Gamma).
\end{equation}
Each Voigt component is defined as
\begin{equation}
V_j(\tilde{\nu}; \tilde{\nu}_{0,j}, A_j, \sigma, \Gamma)
= \frac{A_j}{\sqrt{2\pi}\sigma}
\Re\left[
w\left(
\frac{\tilde{\nu} - \tilde{\nu}_{0,j} + i\Gamma}{\sqrt{2}\sigma}
\right)
\right],
\end{equation}
where $A_j$ and $\tilde{\nu}_{0,j}$ represent the amplitude and central frequency of the $j$-th peak, respectively, while $\sigma$ and $\Gamma$ correspond to the Gaussian and Lorentzian broadening parameters. Here, $w(z)$ denotes the Faddeeva function and $\Re[z]$ indicates the real part of the complex variable $z$. Assuming that the dominant broadening mechanisms are identical for all transitions, the parameters $\sigma$ and $\Gamma$ are constrained to be common to all peaks. The shared-parameters reduce the number of free parameters and improve the robustness and stability of the curve fitting.

From the fitting procedure, we can obtain $\sigma = 0.056(17)~\mathrm{cm^{-1}}$ and $\Gamma = 0.063(20)~\mathrm{cm^{-1}}$, corresponding to an overall FWHM of approximately $0.21(5)~\mathrm{cm^{-1}}$. This spectral resolution is comparable to the typical linewidth of the pulsed Ti:Sa laser used in the experiment. In addition, the fitted peak centers $\tilde{\nu}_{0,i}$ are listed in Table~\ref{table2} and serve as the primary input for the subsequent determination of molecular constants.
\begin{table*}[htbp]
\centering
\captionsetup{justification=raggedright, singlelinecheck=false}
\caption{Comparison between experimentally observed transition wavenumbers $\tilde{\nu}_{0}$ (cm$^{-1}$) with statistical errors $\tilde{\nu}_{0}$ (cm$^{-1}$) and predicted transition wavenumbers $\tilde{\nu}_{0}'$ (cm$^{-1}$) from molecular parameters obtained by PGOPHER line-fitting. The residuals $\Delta \tilde{\nu}_{0}$ are calculated as the observed values minus the predicted ones, while the normalized residuals are calculated as $\Delta \tilde{\nu}_{0}/\sigma_{\tilde{\nu}_{0}}$.}
\label{table2}
\setlength{\tabcolsep}{7pt}
\setlength{\extrarowheight}{3pt}
\begin{tabular}{ccccc}
\toprule
Transition & Observed Wavenumber & Predicted Wavenumber  & Residual & Normalized Residual\\
$\Delta J_{a,b}(J'')$ & $\tilde{\nu}_{0}$ (cm$^{-1}$) & $\tilde{\nu}_{0}'$ (cm$^{-1}$) & $\Delta \tilde{\nu}_{0} = \tilde{\nu}_{0} - \tilde{\nu}_{0}'$ (cm$^{-1}$) & $\Delta \tilde{\nu}_{0}/\sigma_{\tilde{\nu}_{0}}$\\

\midrule
$P_2(85.5)$ & 12176.761(11) & 12176.75(30) & 0.007 & 0.64\\
$P_2(84.5)$ & 12178.110(9) & 12178.12(29) & -0.008 & -0.89\\
$P_2(83.5)$ & 12179.476(5) & 12179.47(28) & 0.003 & 0.60\\
$P_2(82.5)$ & 12180.802(11) & 12180.82(27) & -0.016 & -1.45\\
$P_2(81.5)$ & 12182.169(15) & 12182.16(26) & 0.014 & 0.93\\
\bottomrule
\end{tabular}
\end{table*}

To extract molecular parameters from the observed transitions, we adopt state-specific effective Hamiltonians and perform a standard line-fitting procedure using PGOPHER~\cite{WESTERN2017221}. For the ground state $\mathrm{X}~^2\Sigma^+$, we employ an effective Hamiltonian within Hund's case (b) approximation~\cite{Brown_Carrington_2003}
\begin{equation}
\hat{H}=B\boldsymbol{N}^2-D\boldsymbol{N}^4+\gamma\boldsymbol{N}\cdot\boldsymbol{S}
\end{equation}
where $B$ and $D$ are the rotational constant and centrifugal distortion constant respectively, and $\gamma$ is the spin-rotation coupling constant. For the excited state $\mathrm{A}~^2\Pi$, we use an effective Hamiltonian under Hund's case (a) approximation~\cite{Brown_Carrington_2003}
\begin{equation}
\begin{aligned}
 & \hat{H}=B\boldsymbol{R}^2-D\boldsymbol{R}^4+A\boldsymbol{L}_z\cdot\boldsymbol{S}_z \\
 & -p\frac{1}{2}\left(\boldsymbol{N}_+\boldsymbol{S}_+e^{-2i\phi}+\boldsymbol{N}_-\boldsymbol{S}_-e^{+2i\phi}\right) \\
 & +q\frac{1}{2}\left(\boldsymbol{N}_+^2e^{-2i\phi}+\boldsymbol{N}_-^2e^{+2i\phi}\right)
\end{aligned}
\end{equation}
where $B$ and $D$ again denote the rotational and centrifugal distortion constants, $A$ is the spin–orbit coupling constant, and $p$ and $q$ are the $\Lambda$-doubling constants. The parameter $\phi$ represents the azimuthal angle describing the precession of the electron orbital angular momentum around the internuclear axis. The operators $\boldsymbol{N}_{\pm}$ and $\boldsymbol{S}_{\pm}$ are the ladder operators corresponding to the rotational and electron spin angular momenta, respectively. These effective Hamiltonians form the theoretical basis of spectral assignment and enable a global fit of the molecular constants to the measured transition frequencies.
\begin{table}[htbp]
\centering
\captionsetup{justification=raggedright, singlelinecheck=false}
\caption{Molecular parameters $\mathrm{^{138}BaF}$ obtained in this work, with a comparison with literature value. All quantities are given in units of $\mathrm{cm^{-1}}$.}
\label{table3}
\setlength{\tabcolsep}{7pt}
\setlength{\extrarowheight}{3pt}
\begin{tabular}{cccc}
\toprule
Parameter & this work  &  literature \\
\midrule
\multicolumn{3}{c}{$\mathrm{X}~^2\Sigma^+~(\nu=0)$ } \\
\midrule
$T$ & 234.2443~(fixed) & 234.2443(18)\footnotemark[1] \\
$B$ & 0.215947871~(fixed) & 0.215947871(28)\footnotemark[2] \\
$10^7D$ & 1.84299~(fixed) & 1.84299(23)\footnotemark[3] \\
$10^3\gamma$ & 2.69905~(fixed) & 2.69905(12)\footnotemark[4] \\
\midrule
\multicolumn{3}{c}{$\mathrm{A}~^2\Pi~(\nu=0)$} \\
\midrule
$T$ & 12180.5614~(fixed) & 12180.5614(25)\footnotemark[1] \\
$A$ & 632.2818~(fixed) & 632.2818(27)\footnotemark[5]\\
$B$ & 0.21174(3)& 0.21171(10)\footnotemark[2] \\
$10^7D$ & 2.01(4) & 2.00\footnotemark[3] \\
$p$ & -0.257039~(fixed) & -0.257039(83)\footnotemark[6] \\
$10^3q$ & -0.084~(fixed) & -0.0840(29)\footnotemark[6] \\
\bottomrule
\end{tabular}
\footnotetext[1]{Calculated from values in Ref.\cite{PhysRevA.108.062812} using the formula $T = T_e+\omega_e(\nu+\frac{1}{2})-\omega_ex_e(\nu+\frac{1}{2})^2+\omega_ey_e(\nu+\frac{1}{2})^3$}
\footnotetext[2]{Calculated from values in Ref.\cite{PhysRevA.108.062812} using the formula $B=B_e- \alpha_e (\nu + 1/2)$}
\footnotetext[3]{From Ref.\cite{PhysRevA.108.062812} with a assumption of  $D \approx D_e$}
\footnotetext[4]{From Ref.\cite{PhysRevA.108.062812} with a assumption of  $\gamma \approx \gamma_e$}
\footnotetext[5]{Calculated from values in Ref.\cite{PhysRevA.108.062812} using the formula $A=A_e+ \alpha_A (\nu + 1/2)$}
\footnotetext[6]{From Ref.\cite{Effantin10081990}}
\end{table}
In the line-fitting procedure, all parameters for the ground state are fixed to the high-precision literature values~\cite{PhysRevA.108.062812}. For the excited state, the rotational constant $B$ and the centrifugal distortion constant $D$ are treated as free parameters, while the band origin $T_e$ and spin-orbit constant $A$ are fixed to the high-precision literature values~\cite{PhysRevA.108.062812}. In addition, the $\Lambda$-doublet constants $p,q$ are fixed to reference values~\cite{BERNARD1992174,Effantin10081990}, as no precise information on the $\Lambda$-doubling could be obtained from the measured transitions in this work. The adopted values of fixed parameter, along with the results of the fitted parameters, are summarized in Table.\ref{table3}. The extracted molecular constants $B,D$ for the excited state $\mathrm{A}~^2\Pi_{3/2}$ are highly consistent with the reference value~\cite{PhysRevA.108.062812,BERNARD1992174,Effantin10081990}, both within the $1\sigma$ error bar. Based on the fitted parameters, the predicted spectral lines and their deviations from the observed values in this work are listed in Table.\ref{table2} for a comparison. In general, the deviation on average is approximately less than $0.01~\mathrm{cm^{-1}}$, which is comparable to the typical statistical error of the experimental results. The reduced $\chi^2=1.60$ also indicates that the line-fitting results should be satisfactory. 

For a proof-of-principle experiment, the current resolution and precision are sufficient to validate the feasibility of the technical approach of collinear laser spectroscopy utilizing in-trap produced molecular ion beam. In the future, with further improvements to the spectral resolution, the accuracy of extracted molecular parameter is expected to be enhanced. Currently, the total FWHM of about $0.21~\mathrm{cm^{-1}}$ is still limited by the linewidth of the first-step laser. By switching to an injection-locked Ti:Sa laser system with a narrow linewidth of about 20~MHz, the resolution can be significantly improved~\cite{HU20252721}. This will also enable the extraction of information on the hyperfine structure in molecules such as $\mathrm{^{135,137}BaF}$, arising from the nonzero nuclear spin of the $\mathrm{^{135,137}Ba}$ nucleus.

\section{Summary and Outlook}

In this work, we have successfully and efficiently produced $\mathrm{BaF}^+$ and $\mathrm{YbF}^+$ molecular ions via ion–molecule reactions inside an RFQ-CB. Using the in-trap-produced $\mathrm{BaF}^+$ molecular ion beam and in-flight charge-exchange neutralization, we were able to perform collinear resonance ionization spectroscopy measurements for the $\mathrm{^{138}BaF}$ molecules. By implementing different laser excitation and ionization schemes, we resolved the vibrational and rotational structures of $\mathrm{^{138}BaF}$ involving multiple electronic states. These results demonstrated the feasibility of the integrated methodology that combines molecular formation inside the RFQ-CB with subsequent collinear laser spectroscopy investigation, thereby providing an alternative and promising approach for future spectroscopic studies of radioactive molecules at RIB facilities, such as BRIF in China.

BRIF is capable of producing a variety of radioactive isotopes with 100-MeV proton irradiation of Th- or U-containing targets, providing ample samples for the online in-trap production of radioactive molecules. To date, an online PLASEN system consisting of an RFQ-CB and a collinear resonance ionization spectroscopy setup has been fully constructed and commissioned using short-lived neutron-rich Rb isotopes at BRIF~\cite{PLASEN-online}. Its performance has reached a leading level among similar setups in the worldwide, laying a solid foundation for future online experiments on in-trap production and collinear laser spectroscopy of radioactive molecules.

Additionally, the current in-trap method for molecular ion production can, in principle, be naturally combined with cryogenic buffer gas cooling to generate a cold molecular beam, offering unique advantages in spectroscopic studies of polyatomic radioactive molecules. The development of a cryogenic RFQ-CB for the production of cold radioactive molecules is currently underway at Peking University.

\begin{acknowledgments}
This work was supported by the National Natural Science Foundation of China (12350007, 12305122), New Cornerstone Science Foundation through the XPLORER PRIZE, National Key R\&D Program of China (2023YFA1606403, 2022YFA1605100).
\end{acknowledgments}

\bibliography{Reference}

\end{document}